\documentstyle[pra,aps,psfig,eqsecnum,twocolumn]{revtex}
\begin{document}
\title{\Large \bf Implementing quantum logic operations, 
pseudo-pure states and the Deutsch-Jozsa algorithm using 
non-commuting selective pulses in NMR} 
\author{Kavita Dorai~\cite{k-email}}
\address{Department of Physics, Indian Institute of Science, 
Bangalore 560012 India} 
\author{Arvind~\cite{a-email}}
\address{Department of Physics, Guru Nanak Dev 
University, Amritsar 143005 India}
\author{Anil Kumar~\cite{ak-email}}
\address{Department of Physics and Sophisticated 
Instruments Facility, 
Indian Institute of Science,
Bangalore 560012 India}
\maketitle
\draft
\begin{abstract}
We demonstrate experimentally the usefulness of
selective pulses in NMR to perform quantum
computation.  Three different techniques based on
selective pulse excitations have been proposed to
prepare a spin system in a pseudo-pure state.  We
describe the design of novel ``portmanteau''
gates using the selective manipulation of level
populations.  A selective pulse implementation
of the Deutsch-Jozsa algorithm for a two-qubit
and a three-qubit quantum computer is
demonstrated.
\end{abstract}
\section{Introduction} 
The idea of exploiting
the intrinsically quantum mechanical nature of
physical systems to perform computations has
generated a lot of excitement
recently~\cite{chuang-science-95,divincenzo-science-95}.
Logical operations in quantum computation are
implemented on \underline{qu}antum
\underline{bits}(qubits), where
a qubit can be any quantum two-level system.  The
two eigenstates are mapped onto logical $0\/$ and
$1\/$. While all classical computation can be
performed using the above mapping, the fact that
a qubit can exist in a general coherent
superposition of logical states $0\/$ and $\/1$,
leads to new possibilities for computation.  The
realisation that computation can be performed
reversibly~\cite{feynmann-82,deutsch-royal-89},
paved the way for the quantum mechanical
implementation of logic gates through unitary
transformations~\cite{barenco-pra-95,divincenzo-pra-95}.
The power of quantum computing lies in the fact
that a single input state of a quantum computer
can be a coherent superposition of all possible
classical inputs.  Consequently, algorithms that
are intrinsically quantum in nature can be
designed, to solve problems hitherto deemed
intractable on classical
computers~\cite{shor-siam-97,deutsch-royal-92,grover-prl-97}.
A major hurdle in achieving quantum computing
experimentally, is that of preserving quantum
coherence while the computation is being
performed, and the search for such ideal quantum
systems yielded single charged ions confined in
an ion trap~\cite{cirac-prl-95} and nuclear spins
in a liquid as possible choices.

It has been demonstrated that assemblies of 
nuclear spins in a liquid, which
are largely isolated from their environment and
have long relaxation times(so coherence is
retained for a while), can be used to build
quantum information processors\cite{chuang-science-95}. A system of
$N\/$ spins can exist in entangled quantum
superposition states and can be thought of as an
$N\/$ bit quantum computer. 
However, quantum computing requires pure 
states as inputs, whereas nuclear spins at
thermal equilibrium are in a statistical mixture
of pure states.  It was demonstrated recently that it
is possible to perform quantum computing with
mixed state ensembles rather than on an isolated
system in a pure
state~\cite{cory-proc}. The problem
is circumvented by creating within the overall
density matrix of the system, a sub-ensemble that
behaves like a pure state.  Techniques to prepare
such ``pseudo-pure'' states have been proposed by
different
groups~\cite{cory-proc,knill-quantph,gershenfeld-science-97}.
Previous workers in the field have employed
various NMR methods like non-selective pulses, rf
gradients, coherence transfer via J-coupling and
simultaneous multi-site excitation to create pseudo pure states,
construct universal logic gates and implement quantum
algorithms for two and three-qubit
systems~\cite{ch-nat,ch-prl,ch-roy,j-jcp,j-jmr,lind,madi,vandersypen-qph}.

In this paper, we explore the utility of 
transition-selective and spin-selective pulses, 
and exploit the non-commuting nature of operations on
connected transitions to prepare a spin system in a pseudo-pure
state, execute different logical operations simultaneously and
implement the Deutsch-Jozsa quantum algorithm on a two and a
three-qubit system. The $T_{1}\/$ relaxation times in the
molecules used, is of the order of a few seconds ($3.4\/$ secs - 
$4.6\/$ secs), whereas $T_{2}\/$ relaxation occurs within an
interval of around $1\/$ sec. Selective excitation has been
achieved using low power, long duration pulses of a 
rectangular shape. The length of these pulses is tailored to
achieve sufficient selectivity in the frequency domain 
without perturbing the nearest line, and hence depends on the
magnitude of the smallest $J\/$ coupling present. The
duration of the pulses applied varies from $100\/$ ms to
$263\/$ ms (for $J\/$ couplings of $9.55\/$ Hz to
$3.8\/$ Hz). For small computations, such as the ones
performed here, drastic decoherence or dephasing does
not occur during the duration of these selective pulses. 
However, the deleterious effects of such selective
pulses must be considered and compensated for, whenever
larger computations are attempted.
\section{Creation of pseudo-pure states}
\begin{figure}
\hspace*{1cm}
\psfig{figure=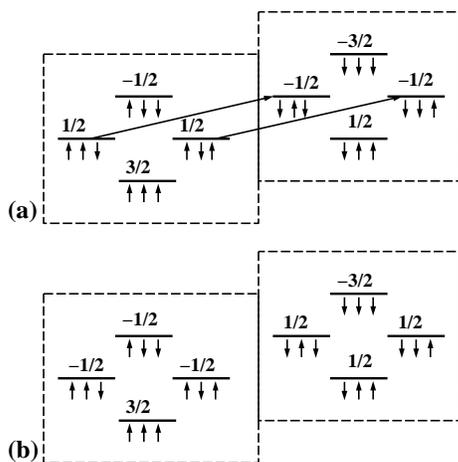,height=6cm,width=6cm,angle=0}
\caption{The creation of a pseudo-pure state in an AMX 
three spin system using
logical labeling. 
(a) The population distribution of the thermal equilibrium state.
(b) The population distribution of a pseudo-pure state, 
created by inverting the populations of the two single-quantum A
transitions shown in (a) by long arrows.}
\label{amx-ppure}
\end{figure}

\begin{figure}
\hspace*{1cm}
\psfig{figure=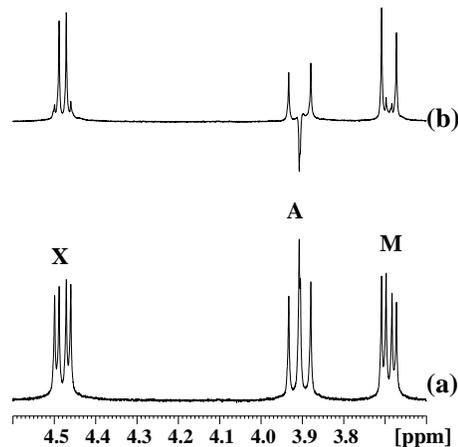,height=6cm,width=6cm,angle=0}
\caption{A logically labeled pseudo-pure state in the 
homonuclear three spin system(AMX) of 2,3 dibromopropionic
acid.
(a) The equilibrium proton spectrum is shown, with
the three protons labeled A, M and X resonating
at  $\delta_{A}=3.91\/$ ppm, $\delta_{M}=3.69\/$ ppm and
$\delta_{X}=4.48\/$ ppm  respectively. 
(b) The selective inversion of the two (nearly overlapping)
central transitions of the A spin leads to the creation of
a logically labeled pseudo-pure state, which has been
read by a small flip angle($10^{0}\/$) detection pulse.
Long, low-power rectangular pulses have been used for
selective excitation.} 
\label{hpure-expt}
\end{figure}

The logical labeling technique to create
pseudo-pure states is broadly categorised by the
fact that unitary transformations are used to
redistribute the populations of states, such that
an effective pure state is obtained in the
sub-manifold of qubits(spins) to be used for
computation, and ancillary qubits are used as
``labels''.  While the concept underlying the
logical labeling method of pseudo-pure state
creation has been delineated by Chuang et.
al.~\cite{gershenfeld-science-97,ch-roy}, there
have been very few experimental implementations
of such an elegant technique~\cite{vandersypen-qph}. We 
have designed a
few novel pulse schemes using transition-selective pulses
to create such logically
labeled pseudo-pure states.  Consider a three
spin-$1/2\/$ system(AMX), with the energy levels labeled
as in Figure~\ref{amx-ppure}.
The selective
inversion of two unconnected single-quantum
transitions of the A spin($\vert \uparrow
\uparrow \downarrow \rangle \rightarrow
\vert \downarrow \uparrow \downarrow \rangle\/$ and
$\vert \uparrow \downarrow \uparrow \rangle \rightarrow
\vert \downarrow \downarrow \uparrow \rangle\/$) would lead to the
creation of a logically labeled pseudo-pure state, with
A being the ``label qubit'' and M,
X being the ``work qubits'' available for computation.
The first four
eigenstates(labeled by the
first spin being in the $\vert\uparrow\rangle\/$ state) 
now form a manifold that
corresponds to a two-qubit pseudo-pure state
while the last four(labeled by the
first spin being in the $\vert \downarrow \rangle\/$ state) 
form
a separate manifold that corresponds to another
two-qubit pseudo-pure state.
The creation of a pseudo-pure state by this method leads
to relative population differences of
\begin{eqnarray}
\begin{array}{cccccccc}
\uparrow\uparrow\uparrow & \uparrow\uparrow\downarrow & 
\uparrow\downarrow\uparrow & 
\uparrow\downarrow\downarrow & 
\downarrow\uparrow\uparrow & \downarrow\uparrow\downarrow & 
\downarrow\downarrow\uparrow & \downarrow\downarrow\downarrow \\
3/2 & -1/2 & -1/2 & -1/2 & 1/2 & 1/2 & 1/2 & -3/2 
\end{array}
\label{sq-pure}
\end{eqnarray}
\noindent{\bf Homonuclear three spin case:}
The experimental creation
of a logically labeled pseudo-pure state in the homonuclear
three-spin system of 2,3 dibromopropionic acid
is shown in Figure~\ref{hpure-expt}.
The pseudo-pure state has
been distilled by manipulating unconnected single quantum
transitions of the label qubit A, as detailed in
Equation~(\ref{sq-pure}). The transition-selective
$\pi\/$ pulses were applied on the two central (nearly overlapping)
transitions of the A spin.

\begin{figure}
\hspace*{1cm}
\psfig{figure=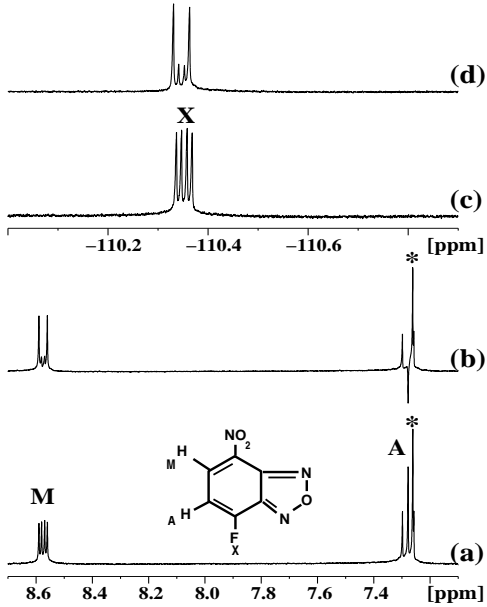,height=10cm,width=7.5cm,angle=0}
\caption{The creation of a logically labeled pseudo-pure state
in 4-fluoro,7-nitro benzofurazan(AMX system).
Transition selective $\pi\/$ pulses 
have been applied on the two
overlapping central transitions of the ``label'' qubit(the proton A). 
The asterisk labels the solvent peak. The normal
proton and fluorine spectra are shown in (a) and (c) with
the coupling constants $J_{AM}=8.1\/$Hz, $J_{AX}=8.01\/$Hz
and $J_{MX}=3.81\/$Hz 
and the spins resonating at
$\delta_{A}=7.28\/$ppm, $\delta_{M}=8.57\/$ ppm and
$\delta_{X}=-110.35\/$ ppm respectively.
The proton and fluorine spectra corresponding to the
pseudo-pure state are shown in (b) and (d) respectively, with
the ``work'' qubits M and X being in a
logically labeled pseudo-pure state.
The state of the system has been monitored by a small
flip angle detection pulse.}
\label{ppure-expt}
\end{figure}

\noindent{\bf Heteronuclear three spin case:}
The experimental creation of a logically labeled pseudo-pure state 
in the heteronuclear three spin system of
4-fluoro,7-nitro benzofurazan is shown in Figure~\ref{ppure-expt}.
Two selective $\pi\/$ pulses were applied on the
central, nearly overlapping 
unconnected single-quantum transitions
($\vert \uparrow \uparrow \downarrow \rangle \rightarrow
\vert \downarrow \uparrow \downarrow \rangle\/$ and
$\vert \uparrow \downarrow \uparrow \rangle \rightarrow
\vert \downarrow \downarrow \uparrow \rangle\/$) of the
A spin(the proton in this case).
The A spin is the ``label'' qubit and the 
other two spins (the third spin being
fluorine in this case) are the ``work'' qubits.
It is to be noted that the spectral pattern of the X spin in the
two pseudo-pure states created in the homonuclear and the
heteronuclear systems are mirror images of each other i.e.
intensities of (0,2,2,0) are obtained for the X spin
in Figure~\ref{hpure-expt}(b) while the X spin multiplet
pattern is (2,0,0,2) in Figure~\ref{ppure-expt}. This
difference reflects
the relative sign of the coupling constants in these
systems.
Heteronuclear ${}^{19}\/$F-${}^{1}\/$H spin systems 
are useful for quantum
computing as they have the twin advantages of good sensitivity
and long relaxation times.

Other  methods to implement a logically
labeled pseudo-pure state can be designed, based 
on the selective
manipulation of the populations of multiple-quanta.
For instance, the inversion of the
double quantum ($\vert\downarrow\uparrow\uparrow\rangle
\rightarrow\vert\downarrow 
\downarrow\downarrow\rangle\/$), followed
by the inversion of the single quantum transition
($\vert \uparrow \downarrow \downarrow \rangle \rightarrow \vert
\downarrow \downarrow \downarrow \rangle\/$), leads
to another pseudo-pure state.
The redistribution of equilibrium populations leads to
relative population differences for the pseudo-pure state
\begin{eqnarray}
\begin{array}{cccccccc}
\uparrow\uparrow\uparrow & \uparrow\uparrow\downarrow &
\uparrow\downarrow\uparrow &
\uparrow\downarrow\downarrow &
\downarrow\uparrow\uparrow & \downarrow\uparrow\downarrow &
\downarrow\downarrow\uparrow & \downarrow\downarrow\downarrow \\
3/2 & 1/2 & 1/2 & 1/2 & -3/2 & -1/2 & -1/2 & -1/2
\end{array}
\end{eqnarray}
Experimentally, the double quantum can be inverted by
a cascade of $\pi\/$ pulses on progressively connected 
single quantum transitions~\cite{kavita-jmr-95}.
The inversion of the single quantum transition
$\vert\downarrow\uparrow\uparrow\rangle\rightarrow 
\vert \downarrow \uparrow \downarrow \rangle\/$,  followed
by the inversion of the zero quantum
$\vert \uparrow \downarrow \downarrow \rangle
\rightarrow \vert \downarrow \uparrow \downarrow \rangle\/$,
leads to yet another pseudo-pure state.
The redistribution of equilibrium populations leads to
the relative population differences
\begin{eqnarray}
\begin{array}{cccccccc}
\uparrow\uparrow\uparrow & \uparrow\uparrow\downarrow &
\uparrow\downarrow\uparrow &
\uparrow\downarrow\downarrow &
\downarrow\uparrow\uparrow & \downarrow\uparrow\downarrow &
\downarrow\downarrow\uparrow & \downarrow\downarrow\downarrow \\
3/2 & 1/2 & 1/2 & 1/2 & -1/2 & -1/2 & -1/2 & -3/2
\end{array}
\end{eqnarray}
Experimentally, the zero quantum can be inverted by
a cascade of $\pi\/$ pulses on two
regressively connected single quantum
transitions by the cascade
$\pi^{1,2}\,\pi^{1,3}\,\pi^{1,2}\/$(spectra not shown)~\cite{kavita-jmr-95}.

The state of the system after the creation of the logically
labeled pseudo-pure state has been read out by a small
flip angle detection pulse in each case. While it is usual
in quantum computing to use pulses of flip angle $90^{0}\/$
for the read out operation, this will not provide a useful
output for logical labeling experiments. To illustrate this
point, consider a logically labeled pseudo-pure state
for the three spin AMX system, with A being the label qubit, 
and M and X the work qubits (Eqn.~\ref{sq-pure}). The 
traceless, deviation density matrix corresponding to this
pseudo-pure state can be described in terms of product
operators (which are just products of spin angular
momentum operators)
\begin{equation}
\sigma_{p-pure} = M_{z} + X_{z} + 4 A_{z} M_{z} X_{z}
\label{flipangle}
\end{equation}
A detection pulse of flip angle $\alpha\/$ leads to
the NMR observable terms
\begin{eqnarray}
&&(M_{x} + X_{x}) \sin{\alpha} + 
(4A_{x}M_{z}X_{z} + 4 A_{z} M_{x} X_{z} +
\nonumber \\
&&
\quad\quad
\quad\quad
\quad\quad
\quad\quad
\quad\quad\quad4A_{z}M_{z}X_{x}) \cos^{2}{\alpha} \sin{\alpha}
\end{eqnarray}
Hence, a $90^{0}\/$ detection pulse would not be able to
read out all the product operators present in the
density matrix (Eqn.~\ref{flipangle}), and a small angle
read pulse is required. 
\section{Novel quantum logic gates}
We now proceed towards the implementation of
quantum logic gates using NMR~\cite{cory-proc,j-jmr}.
The two-qubit quantum XOR(or controlled-NOT) gate has been
demonstrated to be fundamental for
quantum computation~\cite{barenco-pra-95} and 
has been implemented in NMR by a selective $[\pi]_{x}\/$ 
pulse on a single transition~\cite{cory-proc}. 
It has been proved that the reversible quantum XOR gate,
supplemented by a set of general one-qubit quantum
gates, is sufficient to perform any arbitrary quantum
computation~\cite{barenco-pra-95}. Hence, with the 
achievement of the experimental implementation of the XOR 
and one-qubit gates in NMR, it does not seem necessary to look for
the design and construction of other gates.
Nevertheless, it is important in terms of 
complexity in large circuits and ease of experimental
implementation, to look towards the 
design of efficient logic networks.  In this direction, gates
that achieve the implementation of two or more
logic operations simultaneously would be
useful in reducing computational time in circuits
that require a large number of logical operations. We
detail the design and experimental implementation
of such gates here;
borrowing from Lewis Carroll, we call such ``many-in-one''
gates ``Portmanteau'' gates~\cite{lewis-carroll}.

\noindent{\bf The logical SWAP operation:}
Consider a two-spin system (AX) with each spin being a
qubit, the spin A being the first qubit and the spin
X, the second qubit.
The eigen states  of this system can be
represented by 
$\vert \epsilon_{1},\epsilon_{2} \rangle\/$, where
$\epsilon_{1}\/$ and $\epsilon_{2}\/$ are
$0\/$ or $1\/$.
The logical SWAP operation completely exchanges 
the states of a pair of qubits, from  
$\vert \epsilon_{1},\epsilon_{2} \rangle\/$ to
$\vert \epsilon_{2},\epsilon_{1} \rangle\/$,
the unitary transformation corresponding to such an
operation being
\begin{equation}
U_{\mbox{\tiny SWAP}}=\left[\begin{array}{cccc}
1&0&0&0\\
0&0&1&0\\
0&1&0&0\\
0&0&0&1\\ 
\end{array}\right]
\end{equation}
This gate  might be useful during the course of
a computation when qubits need to
be permuted~\cite{j-jmr}.
In spin systems where some scalar J couplings are
ill resolved, the logical SWAP could be used to
compensate for the missing couplings~\cite{madi}. 

Madi et.~al.~\cite{madi}, have implemented the
SWAP operation using an INEPT-type sequence, with
non-selective rf pulses and J-evolution.
It is interesting to note that the logical SWAP operation 
can be achieved by selectively interchanging the populations of the
zero quantum levels.  Since these levels are not connected
by single-quantum transitions, the population exchange
will have to be achieved indirectly.

The inversion of the zero quantum (SWAP) requires a cascade of three
selective $\pi\/$ pulses on regressively connected transitions, 
for example, $[\pi]_x^{A_1} [\pi]_x^{X_1} 
[\pi]_x^{A_1}\/$~\cite{kavita-jmr-95}.
\begin{figure}
\hspace*{1cm}
\psfig{figure=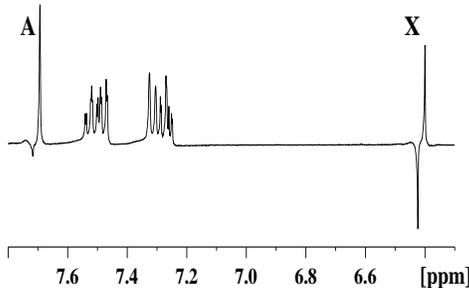,height=4cm,width=6cm,angle=0}
\caption{The logical SWAP operation implemented on the
two-spin system of Coumarin. 
The result of the application of a $[\pi]_x\/$ pulse on one of the
$X\/$ transitions, followed by a $[\pi]_x\/$ pulse on the
regressively connected $A\/$ transition is shown, 
which corresponds to a logical SWAP on the non-equilibrium 
state of the spin system.  A small angle ($10^{0}\/$) 
read pulse has been used.  The 1D spectrum and structure 
of Coumarin is shown in Figure~\ref{cougate}.}
\label{swap}
\end{figure}
Since this would lead to the same spectrum
as the equilibrium spectrum, a non-equilibrium state has been
first created by   
preceding the cascade with a selective $[\pi]_x^{A_1}\/$
pulse, yielding the cascade 
$[\pi]_x^{A_1} [\pi]_x^{X_1} [\pi]_x^{A_1} [\pi]_x^{A_1}=
[\pi]_x^{A_1} [\pi]_x^{X_1}$.
This amounts to the execution of a logical SWAP 
operation on a non-equilibrium state.
The experimental implementation of this operation is
shown in Figure~\ref{swap} on the two spin system of 
Coumarin.

We now explore the implementation of gates that
realise various combinations of 
the SWAP, XOR(XNOR) and NOT operations.
The action and matrix representations of the XOR, XNOR and
NOT gates (all with their outputs on the first qubit)
are given by
\begin{eqnarray}
\vert \epsilon_1,\epsilon_2\rangle 
\stackrel{\rm XOR}{\longrightarrow}
\vert \epsilon_1\oplus\epsilon_2,\epsilon_2 \rangle  
;\quad&
U^{1}_{\mbox{\tiny XOR}} = &\left[\begin{array}{cccc}
1&0&0&0 \\ 
0&0&0&1 \\
0&0&1&0 \\
0&1&0&0 \\
\end{array}
\right]
\nonumber \\
\vert \epsilon_1,\epsilon_2\rangle 
\stackrel{\rm XNOR}{\longrightarrow}
\vert \overline{\epsilon_1\oplus\epsilon_2},\epsilon_2 \rangle  
;\quad&
U^{1}_{\mbox{\tiny XNOR}} = &\left[\begin{array}{cccc}
0&0&1&0 \\ 
0&1&0&0 \\
1&0&0&0 \\
0&0&0&1 \\
\end{array}
\right]
\nonumber \\
\vert \epsilon_1,\epsilon_2\rangle 
\stackrel{\rm NOT}{\longrightarrow}
\vert \overline{\epsilon_1},\epsilon_2 \rangle  
;\quad&
U^{1}_{\mbox{\tiny NOT}} = &\left[\begin{array}{cccc}
0&0&1&0 \\ 
0&0&0&1 \\
1&0&0&0 \\
0&1&0&0 \\
\end{array}
\right]
\end{eqnarray}
The superscript $1$ indicates that the output of the gate is 
obtained on the first qubit. The matrices $U^2_{\mbox{\tiny XOR}}$ 
etc. corresponding to the output on the second qubit can be similarly
constructed.
 
\noindent{\bf Logical SWAP+XOR(XNOR):} The execution of
a logical SWAP operation followed by an XOR gate(with
its output on the first qubit), 
can be defined
through its action on 
$\vert \epsilon_{1},\epsilon_{2} \rangle\/$, 
and leads to the final state 
$\vert \epsilon_{1} \oplus \epsilon_{2},
\epsilon_{1} \rangle\/
$(or to $\vert \overline{ \epsilon_{1} \oplus
\epsilon_{2}}, \epsilon_{1} \rangle\/$ for a SWAP followed
by an XNOR gate with its output on the first qubit)
with their explicit matrix representations being
\begin{eqnarray}
U_{\mbox{\tiny SWAP+XOR}}=
U^{1}_{\mbox{\tiny XOR}}
\,U_{\mbox{\tiny SWAP}}=
\left[\begin{array}{cccc}
1&0&0&0\\
0&0&0&1\\
0&1&0&0\\
0&0&1&0\\ 
\end{array}\right] \nonumber \\
U_{\mbox{\tiny SWAP+XNOR}}=
U^{1}_{\mbox{\tiny XNOR}}
\,U_{\mbox{\tiny SWAP}}=
\left[\begin{array}{cccc}
0&1&0&0\\
0&0&1&0\\
1&0&0&0\\
0&0&0&1\\ 
\end{array}\right]
\end{eqnarray}

An implementation of these operations 
requires the application of selective $[\pi]_{x}\/$ 
pulses consecutively
on two regressively connected transitions, 
and the resulting spectrum on two
bits is identical to Figure~\ref{swap}. 

\noindent{\bf XOR(XNOR)+Logical SWAP+NOT:} The 
implementation of an XOR 
gate(with the output on the first qubit), 
followed by a SWAP operation
and then a  
NOT gate on the first qubit, corresponding to a final
state of $\vert \overline{\epsilon_{2}},
\epsilon_{1} \oplus \epsilon_{2} \rangle \/$
can be experimentally achieved by 
transition-selective $[\pi]_{x}\/$ pulses applied consecutively on
two progressively connected transitions.
Reversing the order of application of the pulses leads to
the final state 
$\vert \overline{\epsilon_{1}\oplus\epsilon_{2}},
\overline{\epsilon_{1}} \rangle \/$ 
which corresponds to an XNOR+Logical SWAP+NOT gate, with 
the output on the second qubit. 
 
These gates correspond to the unitary matrices given by
\begin{eqnarray}
U_{\mbox{\tiny XOR+SWAP+NOT}}=
U^{1}_{\mbox{\tiny NOT}}
\,U_{\mbox{\tiny  SWAP}}
\,U^{1}_{\mbox{\tiny  XOR}}=
\left[\begin{array}{cccc}
0&0&0&1\\
0&1&0&0\\
1&0&0&0\\
0&0&1&0\\ 
\end{array}\right] \nonumber \\
U_{\mbox{\tiny XNOR+SWAP+NOT}}=
U^{2}_{\mbox{\tiny NOT}}
\,U_{\mbox{\tiny SWAP}}
\,U^{2}_{\mbox{\tiny XNOR}}=
\left[\begin{array}{cccc}
0&0&1&0\\
0&1&0&0\\
0&0&0&1\\
1&0&0&0\\ 
\end{array}\right]
\nonumber 
\end{eqnarray}
\vspace*{-12pt}
\begin{equation}
\label{combined-gate}
\end{equation}

\begin{figure}
\psfig{figure=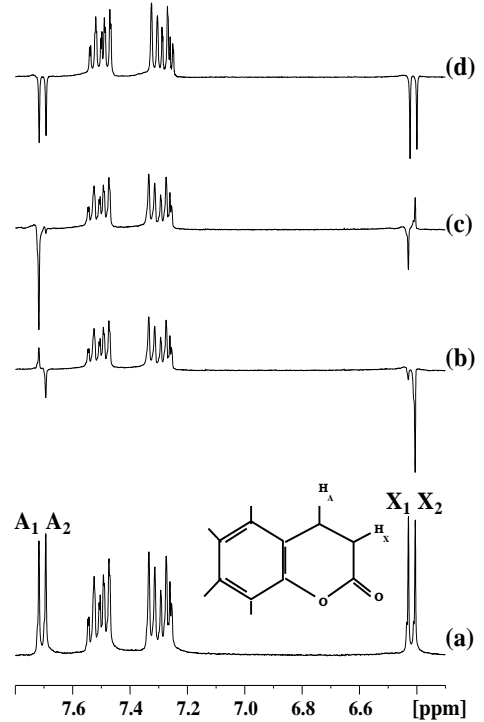,height=12cm,width=8cm,angle=0}
\caption{The experimental implementation of various portmanteau gates.
(a) Reference spectrum of Coumarin at thermal equilibrium.
(b) The implementation of an 
XOR + SWAP + NOT (by the pulse cascade $[\pi]_x^{A_1}$
followed by 
$[\pi]_{x}^{X_2}$, where 
$A_{1}\/$ and $X_{2}\/$ 
refer to progressively connected transitions of
spins A and X respectively).
(c) The implementation of an XNOR + SWAP + 
NOT 
($[\pi]_x^{X_2}$ followed by $[\pi]_x^{A_1}\/$). 
(d) The implementation of a NOT + SWAP 
($[\pi]_x^{A_1} 
[\pi]_x^{X_2} [\pi]_x^{A_1}\/$).
The state of the spin system is read by a small ($10^{0}\/$) angle
pulse in each case.}
\label{cougate}
\end{figure}

The experimental implementation of such gates is 
shown in~Figure~\ref{cougate}(b),(c).
It is interesting to note that these operations do not
commute, so the order in which the pulses are applied is 
important and its reversal leads to
different logical operations.

\noindent{\bf NOT+Logical SWAP:} 
The NOT 
gate  followed by  a  logical SWAP operation on two 
qubits (or vice versa since these operations commute) 
leads to 
$\vert \overline{\epsilon_2}, 
\overline{\epsilon_1}\rangle\/$ when applied to
the state $\vert \epsilon_1,\epsilon_2 \rangle\/$.
This action suffices to determine the unitary matrix 
for the above  operation, given by
\begin{equation}
U_{\mbox{\tiny NOT+SWAP}}=
U_{\mbox{\tiny SWAP}}
\,U_{\mbox{\tiny NOT}}=
U_{\mbox{\tiny NOT}}
\,U_{\mbox{\tiny SWAP}}=
\left[\begin{array}{cccc}
0&0&0&1\\
0&1&0&0\\
0&0&1&0\\
1&0&0&0\\ 
\end{array}\right]
\end{equation}

The experimental implementation 
has been achieved by 
selectively inverting the 
populations of the double-quantum
levels. A cascade of transition-selective
$[\pi]_{x}\/$ pulses has been applied on two
 progressively connected 
transitions~(Figure~\ref{cougate}(d)).

The implementation of various portmanteau 
gates on a thermal initial state is shown in Figure~\ref{cougate} 
for the two-spin system of Coumarin.
The same pulse schemes are expected to 
implement the  above logic operations on other initial 
states (for instance, a pseudo-pure or a 
coherent superposition of states)  
as well. As an illustration, consider the portmanteau gate
{XOR+SWAP+NOT}. The unitary matrix corresponding to it 
(Eqn.~\ref{combined-gate}) can be decomposed into two matrices
\begin{eqnarray}
&U_{\mbox{\tiny XOR+SWAP+NOT}} =
U_{\pi_x}^{X_{2}} 
U_{\pi_x}^{A_{1}};&
\nonumber \\
&U_{\pi_x}^{A_{1}} =\left[\begin{array}{cccc}
1 & 0 & 0 & 0 \\
0 & 1 & 0 & 0 \\
0 & 0 & 0 & i \\
0 & 0 & i & 0 \\
\end{array}
\right], 
U_{\pi_x}^{X_{2}} = \left[ \begin{array}{cccc}
0 & 0 & i & 0 \\
0 & 1 & 0 & 0 \\
i & 0 & 0 & 0 \\
0 & 0 & 0 & 1 \\
\end{array}
\right]&
\end{eqnarray}
The matrices $U_{\pi_x}^{A_{1}}\/$ and $U_{\pi_x}^{X_{2}}\/$ 
correspond to  selective $[\pi]_{x}\/$ pulses on the $A_{1}\/$ 
and $X_{2}\/$ transitions respectively. A low-power, long 
duration rectangular pulse is applied to achieve the desired 
selectivity. These transition-selective pulses can be 
expanded in terms of single-transition operators (the expansion 
is independent of the state of the spin system~\cite{kavita-jmr-95,ernst}).
This  realisation in terms of single transition operators is 
valid when the power of the rf pulse is low compared to the  
the J-coupling and the chemical shift difference between the spins 
($w_1 << 2\pi J << \delta_{AX}\/$). 
One is thus able to realise the unitary transformations required to 
implement the desired logical operations, without prior knowledge 
of the state of the system.
However as noted recently~\cite{cory-qph}, 
a more complete Hamiltonian description might be required to describe 
evolution under selective pulses,
in order to fully establish the generality of the above schemes. 

It is to be noted that we have used selective pulses polarised
along the $x\/$ direction. 
The phases of the pulses are
important, 
and have been experimentally
ensured by appropriate phase cycling schemes.  
\vspace*{-12pt}
\section{Implementation of the Deutsch-Jozsa(D-J) quantum algorithm}
Finally, we experimentally implement the
D-J algorithm using selective pulses. 
The D-J algorithm determines 
whether an unknown function $f(x)\/$ is
constant or balanced~\cite{deutsch-royal-92}. 
In the simplest version, $f(x)\/$ is a
mapping from a single bit to a single bit and the function is
constant if  $f(x)\/$ is independent of $x\/$ and it is
balanced if $f(x)\/$ is zero for one value of $x\/$ and
unity for the other value. 
The generalisation to $N\/$ bits is
conceptually simple and $f(x)\/$ in this case is constant
if it is independent of $x\/$ and balanced if it is zero for
half the values of $x\/$ and unity for the other half. 
A classical computer proceeding deterministically would
require up to $2^{N-1} + 1\/$ function calls to check if
$f(x)\/$ is constant or balanced; even if half the
inputs have been evaluated and all outputs have been found
0(or 1) one cannot conclude that the function is
constant.  The quantum version of the
algorithm determines if the function is balanced or
constant using only a {\em single} function call. For the 
one bit case this is achieved by evaluating
the value of $f(0) \oplus f(1)\/$
(where $\oplus\/$ denotes addition modulo 2).
The binary function $f\/$ is
encoded in a unitary transformation by the
propagator $U_{f}\/$ by including an extra input qubit such that
$\vert x \rangle \vert y \rangle
\stackrel{U_{f}}{\longrightarrow} \vert x \rangle
\vert y \oplus f(x) \rangle \/$.
The four possible functions
for the single-bit D-J algorithm are categorised as

\begin{center}
\begin{tabular}{|p{6pt}cp{6pt}|p{6pt}cp{6pt}|
 p{6pt}cp{6pt}|p{6pt}cp{6pt}|p{6pt}cp{6pt}|}
\hline
&&& \multicolumn{6}{c|}{CONST.} &
\multicolumn{6}{|c|}{BAL.} \\ 
\hline
&$x$&&&$f_{1}$&&&$f_{2}$ &&& $f_{3}$ &&& $f_{4}$&\\
\hline
&0 &&& 0 &&& 1 &&& 0 &&& 1&\\
&1 &&& 0 &&& 1 &&& 1 &&& 0&\\
\hline
\end{tabular}
\end{center}

The unitary transformations corresponding to the four
possible propagators $U_{f}\/$  
can be easily constructed:
\begin{eqnarray}
U_{1} = \left[\begin{array}{cccc}
1 & 0 & 0 & 0 \\
0 & 1 & 0 & 0 \\
0 & 0 & 1 & 0 \\
0 & 0 & 0 & 1
\end{array}
\right]
U_{2} = \left[\begin{array}{cccc}
0 & 1 & 0 & 0 \\
1 & 0 & 0 & 0 \\
0 & 0 & 0 & 1 \\
0 & 0 & 1 & 0
\end{array}
\right] \nonumber \\
U_{3} = \left[\begin{array}{cccc}
1 & 0 & 0 & 0 \\
0 & 1 & 0 & 0 \\
0 & 0 & 0 & 1 \\
0 & 0 & 1 & 0
\end{array}
\right]
U_{4} = \left[\begin{array}{cccc}
0 & 1 & 0 & 0 \\
1 & 0 & 0 & 0 \\
0 & 0 & 1 & 0 \\
0 & 0 & 0 & 1
\end{array}
\right]
\end{eqnarray}
The algorithm requires one input spin and one work spin.
Using the propagator $U_{f}\/$ and appropriate input states,
one can proceed with the implementation of the algorithm.
Previous workers in the field~\cite{ch-nat,j-jcp},
used a combination of spin-selective $\textstyle (\pi/2)\/$
pulses and evolution under the scalar coupling J, to
encode the D-J algorithm on a two-qubit quantum computer. 

We have implemented the D-J algorithm
using spin-selective and transition-selective
$\pi\/$ pulses. The experiment begins with both qubits 
in a superposition of states, achieved by
a non-selective $[\pi/2]_{y}\/$ pulse on both spins.
After application of the propagators $U_{i}\/$, the first
qubit(the ``control'' qubit) remains in the 
superposition state while the
desired result $(f(0) \oplus f(1))\/$ is encoded as the
appearance or disappearance of the lines of the target qubit.
The $U_{1}\/$ transformation corresponds to the unity operation
or ``do nothing'', while the $U_{2}\/$ transform is achieved by
a spin-selective $[\pi]_{x}\/$  pulse on the control qubit.
The $U_{3}\/$ and $U_{4}\/$ transformations are implemented by
selective $\pi\/$ pulses on the
$\vert \downarrow \uparrow \rangle 
\rightarrow \vert \downarrow \downarrow \rangle\/$
and the
$\vert \uparrow \uparrow \rangle 
\rightarrow \vert \uparrow \downarrow \rangle\/$
transitions respectively.

\begin{figure}
\hspace*{1cm}
\psfig{figure=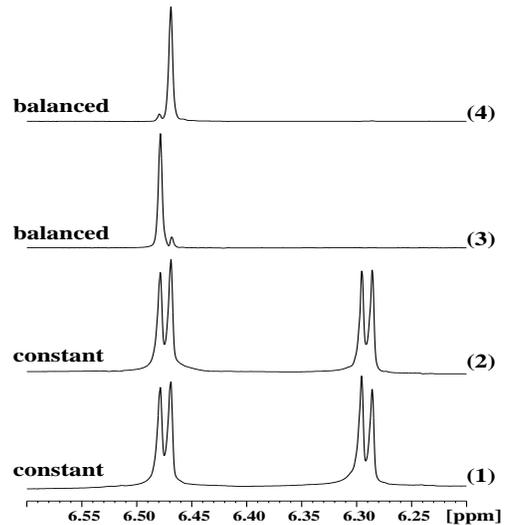,height=7.0cm,width=6.5cm,angle=0}
\caption{A selective pulse implementation of the
D-J quantum algorithm on a two-qubit system
5-nitro furaldehyde, at room temperature on a 400 MHz spectrometer. 
The results after applying the unitary transformations
$U_{1}\/$, $U_{2}\/$, $U_{3}\/$ and $U_{4}\/$ on a coherent
superposition are shown in (1), (2), (3) and (4)
respectively.}
\label{furdj}
\end{figure}
The implementation of this algorithm does not require 
pure initial states as similar results can be
read out from the spectrum if one starts with thermal initial states
instead. In a single measurement, one can distinguish between
constant and balanced functions on the basis of the
disappearance of the lines of the target qubit in the spectrum.
These predictions are borne out by the
experimental spectra in Figure~\ref{furdj}.
The phase of the transition-selective pulse (to implement
$U_{3}\/$ and $U_{4}\/$) has been stepped through
$(x,-x,y,-y)\/$ to suppress phase distortions and leads to
the total suppression of the target qubit lines and the
retention of only one line of the control qubit.

The algorithm to distinguish between the
two categories(constant or balanced) of two-bit binary
functions is implemented on a three-qubit NMR computer, by
evaluating 
$ \vert x \rangle \vert y \rangle \vert z \rangle
\stackrel{U_{f}}{\longrightarrow} \vert x \rangle \vert y \rangle
\vert z \oplus f(x,y) \rangle\/$. The eight possible
(2 constant and 6 balanced) two-bit binary functions are
categorised as

\begin{center}
\begin{tabular}{|p{2pt}cp{2pt}|p{2pt}cp{2pt}|p{2pt}cp{2pt}|
p{2pt}cp{2pt}|p{2pt}cp{2pt}|
p{2pt}cp{2pt}|p{2pt}cp{2pt}|
p{2pt}cp{2pt}|p{2pt}cp{2pt}|
p{2pt}cp{2pt}|p{2pt}cp{2pt}|}
\hline
&&&&&& \multicolumn{6}{c|}{CONST.} &
\multicolumn{18}{c|}{BAL.} \\
\hline
&$x$ &&& $y$ &&& $f_{1}$ &&&  $f_{2}$ &&&
$f_{3}$ &&&  $f_{4}$ &&&  $f_{5}$ &&&
$f_{6}$ &&&  $f_{7}$ &&&  $f_{8}$& \\
\hline
&0 &&& 0 &&& 0 &&& 1 &&& 0 &&& 1 &&& 1 &&& 0 &&& 1 &&& 0& \\
&0 &&& 1 &&& 0 &&& 1 &&& 0 &&& 1 &&& 0 &&& 1 &&& 0 &&& 1& \\
&1 &&& 0 &&& 0 &&& 1 &&& 1 &&& 0 &&& 1 &&& 0 &&& 0 &&& 1& \\
&1 &&& 1 &&& 0 &&& 1 &&& 1 &&& 0 &&& 0 &&& 1 &&& 1 &&& 0& \\
\hline
\end{tabular}
\end{center}  

Previous researchers used shaped pulses generated by
an rf waveform generator to implement the two-bit D-J 
algorithm using three qubits; the pulse waveforms were
tailored to selectively excite two or more frequencies
simultaneously~\cite{lind}.

\begin{figure}
\psfig{figure=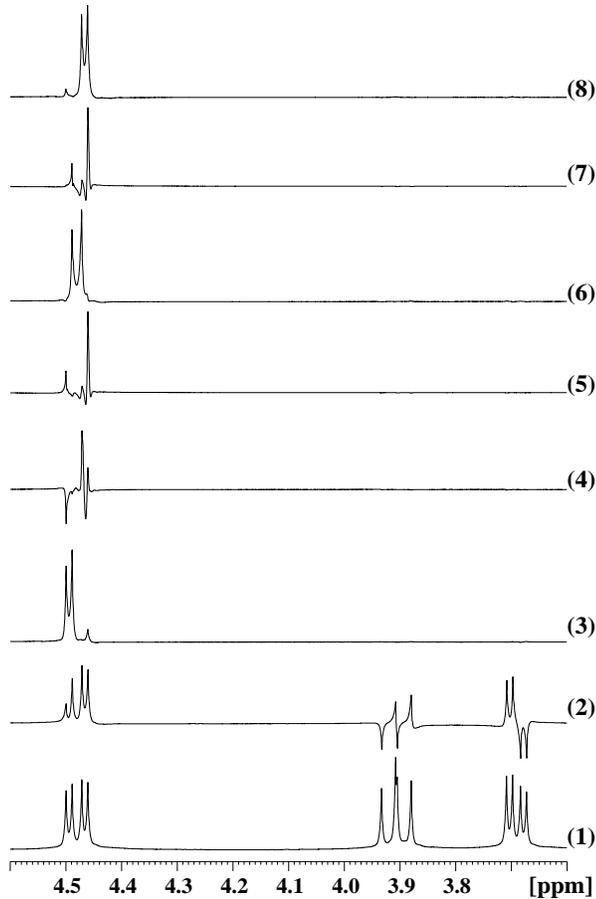,height=12cm,width=8cm,angle=0}
\caption{Selective pulse implementation of the
D-J algorithm on the three-qubit system of
2,3 dibromopropionic acid. The two constant functions $f_{1}\/$
and $f_{2}\/$ are
shown in (1) and (2) respectively, while the unitary transforms
corresponding to the
balanced functions $f_{3}-f_{8}\/$ are implemented in
(3)-(8) respectively.}
\label{dibdj}
\end{figure}

We describe here a selective pulse implementation 
of the D-J algorithm using simple rectangular 
pulses (Figure~\ref{dibdj}). 
These low power, long duration
transition-selective pulses are applied consecutively, and
do not require any special hardware for their application.
The unitary transforms have been implemented on a
coherent superposition of all the three qubits, achieved
by a non-selective $\pi/2\/$ pulse on a thermal initial state.
The two constant functions $f_{1}\/$ and $f_{2}\/$ correspond to
the unity operation and a spin-selective $[\pi]_{x}\/$ pulse on the
multiplet of the control qubit, respectively. The 
unitary transformations encoding the
six balanced functions $f_{3}-f_{8}\/$ are implemented  by
selective pulses on the transitions of the control qubit, taken
two at a time  i.e. the pulses can be described by
$[\pi,\pi,0,0]\/$, $[\pi,0,\pi,0]\/$, $[\pi,0,0,\pi]\/$,
$[0,\pi,\pi,0]\/$, $[0,\pi,0,\pi]\/$ and $[0,0,\pi,\pi]\/$ on the
four transitions where 0 denotes no pulse on that particular transition.
The phases of the transition selective pulses have been
stepped through $(x,-x,y,-y)\/$ as in the two-qubit case and
a similar logic prevails in explaining the spectral pattern
obtained. Unlike the previous implementation of
the three-qubit D-J algorithm~\cite{lind}, the phase cycling here 
achieves a complete suppression of the
multiplets of both the target spins when the function
is balanced ($f_{3} - f_{8}\/$).

It has been demonstrated that selective pulse techniques
in NMR are a powerful tool to build quantum information
processors. The distilling of a pseudo-pure state from
a thermal state, the simultaneous implementation of 
different logical operations to save computational time and
a two and three-qubit implementation of the D-J
quantum algorithm has been experimentally achieved using such
techniques.

\centerline{\Large \bf Acknowledgements}
The use of the AMX-400 high resolution FTNMR spectrometer of
the Sophisticated Instruments Facility, Indian Institute of
Science, Bangalore, funded by the Department of Science
and Technology, New Delhi, is gratefully acknowledged.
\end{document}